\documentclass[reprint,fleqn,superscriptaddress,twocolumn,showpacs,
amsmath,amssymb,prb,aps,bibliography]{revtex4-1}
\usepackage[all]{xy}
\usepackage{gensymb}
\usepackage{graphicx,psfrag,times,epsfig,color}
\usepackage{verbatim,natbib}

\usepackage{color}
\usepackage{makeidx}
\usepackage{amsmath}
\usepackage{bm}
\usepackage{amsfonts}
\usepackage{amssymb}
\usepackage{hyperref}

\begin{document}

\title{Generalization of interlayer tunneling models 
to cuprate superconductors with charge density waves}
\author{H\'ercules Santana and E. V. L. de Mello}
\affiliation{Instituto de F\'{\i}sica, Universidade Federal Fluminense, 24210-346 Niter\'oi, RJ, Brazil}

\email[Corresponding author: ]{evandro@mail.if.uff.br}

\begin{abstract}

At the beginning of cuprate superconductors, the interlayer tunneling (ILT) and Lawrence-Doniach 
(L-D) models, which connect the CuO planes by Josephson coupling, were considered the leading 
theoretical proposals for these materials. 
However, measurements
of the interlayer magnetic penetration depth $\lambda_{c}$ yielded larger values
than required by the ILT model. After the discovery of planar stripes
and incommensurate charge ordering, it was also possible to consider Josephson coupling
between these mesoscopic charge domains or blocks. We show that the average intralayer
is larger than the interlayer coupling and comparable with the condensation energy,
leading to a superconducting transition by long-range phase order. 
Another consequence is that the ratio $[\lambda_{c}/\lambda_{ab}]^2$ 
is related to the resistivity ratio $\rho_c/ \rho_{ab}$ 
near the superconducting transition temperature in agreement with several measurements.

\end{abstract}
\maketitle

\section{Introduction}

Several experiments on high-temperature cuprate
superconductors (HTS) verified their large
anisotropic properties that arise mainly because of the much smaller 
resistivity ($\rho_{\rm ab}$) along with the CuO layers.
These facts suggested that the ILT\cite{Clem89,Clem91,Chakravarty1993,Anderson1995,Leggett1996} 
or L-D models\cite{LD1971}, 
which describes a layered superconductor as a stack of Josephson-coupled 
adjacent blocks or layers were ideal candidates to describe the HTS. The 
ILT model of Anderson and collaborators\cite{Chakravarty1993,Anderson1995} 
consisted of non-Fermi liquid 
planar electrons and, in the superconducting (SC) phase,
interlayer tunneling of Cooper pairs. This approach results in a strong decrease of the 
$c$-axis kinetic energy with a concomitant increase of the condensation energy
(the gain of free energy in the SC state compared with the normal state).

On the experimental side, Shibauchi 
{\it et al}\cite{Shibauchi94} measured the magnetic penetration depth 
$\lambda_{\rm c}$ and the planar $\lambda_{\rm ab}$
in single crystals of La$_{2-x}$Sr$_x$CuO$_4$ (LSCO) and found that $\lambda_{\rm c}(T)$
was in good agreement with the L-D model. However, images of interlayer Josephson vortices\cite{Images1998} 
in single-layer compounds Tl$_2$Ba$_2$CuO$_6$
yielded about 20 $\mu$m which is 
about 20 times the penetration depth determined by the ILT model\cite{Leggett1996}. 
This result was considered a strong 
evidence against the ILT and L-D models to cuprates\cite{Images1998,Anderson1998}
and they were abandoned as the leading general HTS theories. 

On the other hand, over the years, a significant number of new experiments 
with novel techniques and methods 
revealed properties not known when the ILT was originally proposed, which
opened new possibilities: 
In particular, charge inhomogeneities in the form of stripes were discovered in underdoped
Nd substituted in LSCO by neutron
scattering\cite{Tranquada1995a}, which was a key experimet to
the detection of charge-ordering (CO) or charge density waves (CDW)
phenomena in HTS. 
Along this line, scanning tunneling microscopy (STM)
experiments made it possible to obtain atomically resolved 
maps\cite{Lang2002,STM2007} on the energy-dependent local density of states (LDOS). 
More recently, resonant x-ray diffraction (REX) revealed the subtle variations
of the CO wave length $\lambda_{CO}$ with the doping level of several families
of cuprates\cite{Comin2016}.

To interpret their inhomogeneous STM data on underdoped 
Bi$_2$Sr$_2$CaCu$_2$O$_{8+\delta}$ (Bi2212), Lang {\it et al}\cite{Lang2002}
proposed a structure of mesoscopic superconductors grains
connected by Josephson coupling.
This original proposal was not generally accepted, mainly because 
there was no evidence of CO instability near the optimal and in the overdoped region.
However a few years later, similar CO granular patterns were observed by STM near the
optimal value\cite{Wise2008} and even in the overdoped regions
\cite{Gomes2007,Parker2010,He2014}. Furthermore,
a variety of complementary experimental probes detected charge instability 
in all hole-doped HTS
families\cite{Comin2016} as well as in Nd-based electron-doped\cite{Ndoped2018}. 
Recently, charge inhomogeneities have been detected in overdoped LSCO up 
to at least $x = p = 0.21$\cite{Wu2017,Shen2019,Fei2019}
and possibly up to $p = 0.25$\cite{Tranquada2021}. Therefore, the ubiquitous presence
of CDW in all HTS compounds suggested that they are intertwined with the
SC phase and somehow related to the SC interaction\cite{DeMello2012,deMello2014,Mello2017,Mello2020a}.

To understand the way they intertwine we use the Cahn-Hilliard (CH) 
equation that can simulate
the observed CO wavelength $\lambda_{\rm CO}$ of different
materials employing a phase separation Ginzburg-Landau (GL) free energy. 
The $V_{\rm GL}$ free energy can be tuned in different forms or shapes
and acts as a template for the CO or CDW while confining the charges in alternating
hole-rich and hole-poor domains. $V_{\rm GL}$ works like a surface potential that binds
the electrons in physical grains of a granular superconductor with the 
difference that the CDW domains
are of nanoscopic dimensions. But the Cooper pairs coherence lengths in HTS are also of nanoscopic
sizes, and they may be formed by local hole pairs interaction 
mediated by $V_{\rm GL}$ modulations. 

In this scenario, we calculate local SC amplitudes
by a self-consistent Bogoliubov-deGennes (BdG) approach.
Akin to granular superconductors, there are  
Josephson coupling between the nanoscopic charge domains that compete with 
thermal disorder to promote long-range phase order at
the SC critical temperature $T_{\rm c}$\cite{Mello2020a,Mello2022}.
We also consider the planar Josephson coupling 
between the CO domains together with 
interlayer coupling to formulate a generalization of the ILT and L-D models.

We mentioned above that the measurements and calculations of the penetration 
depth $\lambda_c$ were important tests to the ILT theories.
On the other hand, the Josephson couplings are proportional 
to the local superfluid  densities\cite{Spivak1991} that, in turn are 
 proportional to the inverse 
of the magnetic penetration depth\cite{Bozovic2016}, which is our route to estimate 
$\lambda_c$ and $\lambda_{ab}$. Using the LSCO calculations from Ref. \onlinecite{Mello2020a}
we reproduce several low temperatures $(\lambda_{c}(p)/\lambda_{ab}(p))^2$
measurements\cite{Shibauchi94,Panagopoulos2000}.
We also demonstrate a new equation relating this ratio
to the resistivities $R^{c}_{\rm n}/R^{ab}_{\rm n}$ just above the SC transition, 
which is easy to test experimentally and is 
in agreement with several old measurements\cite{Kimura92,Nakamura93}.

\section{CDW Calculations}

We mentioned that the CH phase separation method 
reproduces the observed planar CDW,
but its great advantage is the GL free energy map that provides a scale
to the pairing attraction.
The starting point is the time-dependent
phase separation order parameter associated with the relative local electronic
density, $u({\bf r},t) = (p({\bf r},t) - p)/p$, where 
$p({\bf r},t)$ is the  local charge or hole density at a 
position ${\bf r}$ in the CuO plane. 
The CH equation is based on the
GL free energy expansion in terms of this (conserved) order parameter 
$u$\cite{Otton2005,deMello2009,DeMello2012}: 

\begin{equation}
f(u)= {{\frac{1}{2}\varepsilon |\nabla u|^2 + V_{\rm GL}(u,T)}},
\label{FE}
\end{equation}
where $\varepsilon$ is the parameter that controls the charge modulations
and ${V_{\rm GL}}(u,T)= -\alpha [T_{\rm PS}-T] u^2/2+B^2u^4/4+...$ is a
double-well potential that characterizes the two (hole-rich and hole-poor) local charge
densities of the CDW structure. The phase separation transition 
temperature $T_{\rm PS}$ is 
assumed to be near the pseudogap instability at $T^*(p)$. 

An elegant way to derive the CH equation is through the continuity equation 
for the local free energy current density ${\bf J}=M{\bf\nabla}(\delta f/ \delta u)$,\cite{Bray1994}
\begin{eqnarray}
\frac{\partial u}{\partial t} & = & -{\bf \nabla.J} \nonumber \\
& = & -M\nabla^2[\varepsilon^2\nabla^2u
- \alpha^2(T)u+B^2u^3].
\label{CH}
\end{eqnarray}

The equation is non-linear and solved by a stable and fast finite difference
scheme with free boundary conditions, and we stop the simulation time
$t$ when a given CDW structure is reproduced and the solution $u({\bf r})$
or $p({\bf r})$ are used in the SC calculations.

\begin{figure}% figure* is for twocolumn figures.
%\begin{center}
%\centering
%\begin{minipage}{width=0.46\textwidth}
\centerline{\includegraphics[height=4.0cm]{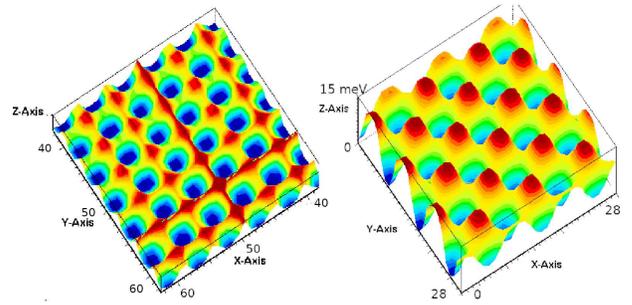}}
%end{center}
%i\end{minipage}
\caption{ a) A three-dimensional plot of the two-dimensional phase separation potential 
${V_{\rm GL}}({\bf r}, T)$ viewed from just
above the CuO plane. Notice the array of similar potential wells that host rich and poor alternating 
charge density domains $p({\bf r})$. 
b) Similar view of the SC amplitudes $\Delta_d({\bf r})$ that also follows the same modulation pattern, 
what is known as pair density waves. 
}
\label{fig1}
\end{figure}

We have provided a detailed description of the CH simulations in several previous 
works\cite{Mello2020a,Mello2021,Mello2022}. 
Figure \ref{fig1}(a) illustrates a typical 
{\color{red}
${V_{\rm GL}}({\bf r}, T) \equiv {V_{\rm GL}}(u({\bf r}), T)$ }
low temperature solution for a $p = 0.19$ LSCO compound.  We can see that
${V_{\rm GL}}({\bf r}, T)$ form an array of side-by-side potential minima that
hosts the alternating hole-rich and hole-poor charge densities domains
(not shown here, see many simulations in the supplemental material of Ref. \onlinecite{Mello2022}).

The derived CDW density map that reproduces the measurements of a given compound and the
respective ${V_{\rm GL}}$ will be used in the BdG calculations to obtain the SC 
properties in the next section.

\section{The BdG superconducting Calculations}

To perform the BdG SC approach, we use two results from the CH calculations:\\

1- The CDW density map $p({\bf r})$.\\

2- The functional $V_{\rm GL}({\bf r})$ shown in Fig. \ref{fig1}(a)\\

As mentioned in the introduction, at low temperatures, 
$V_{\rm GL}({\bf r}, T)$ constrains the planar charges 
in alternating hole-rich and hole-poor domains forming the CDW structure. 
These alternating densities force the ions
to oscillate around new displaced positions as observed by x-ray diffraction\cite{Chang2012}.
They interact back with the holes,
leading to a local lattice-mediated hole-hole attraction 
{\color{red} 
and Cooper pairs inside the CDW domains, recalling that this possible because the SC coherence
length $\xi$ is in general shorter than $\lambda_{CO}$.}
This pairing interaction is dependent
on the CDW hole-rich and hole-poor local concentrations, and it is reasonable
to assume that it scales with 
the localization potential $V_{\rm GL}({\bf r}, T)$\cite{Mello2020a}.

We use this interaction as a nearest
neighbor potential attraction in an extended Hubbard model to calculate 
the local SC amplitudes $\Delta_d({\bf r}_i)$. This is done by
a self-consistent approach that keeps the $p({\bf r}_i)$ CDW structure constant from the beginning
to the end of the calculations following several different
experiments\cite{Wise2008,Gomes2007,Parker2010,Chang2012,Tranquada2021}. This is achieved by changing the local chemical potential $\mu({\bf r}_i)$ at each iteration until 
the $\Delta_d({\bf r}_i)$ amplitudes converge and the density map $p({\bf r}_i)$
is preserved. 
At the end of the calculations we obtain the original CDW map and
the local $d$-wave amplitudes $\Delta_d({\bf r}_i)$ with the same charge modulations   
($\lambda_{\rm CO}(p)$) what is known as pair density waves (PDW)\cite{Mello2017}. 
This is shown in
Fig. \ref{fig1}(b) for the same compound of Fig. \ref{fig1}(a).

The $\Delta_d({\bf r}_i,p, T)$ local spatial variations imply that global properties like
the condensation energies, critical temperatures, and inter-intralayer Josephson
coupling, are a function of the average SC amplitudes\cite{Mello2020a,Mello2021,Mello2022}
given by:
\begin{equation}
{\left  <\Delta_d(p,T)\right >} = \sum_i^N \Delta_d({\bf r}_i,p, T)/N, 
\label{DdM}
\end{equation}
where the sum, like in the case of ${\left < V_{\rm GL}(p)\right >}$, is over the N unit cells 
of a single CuO plane. 

\section{Josephson Coupling Calculations}

We mentioned in the introduction that the CDW structure shown in Fig. \ref{fig1}(a) 
with its charge domains bounded by the $V_{\rm GL} ({\bf r}_i,p, T)$
potential has some similarities with granular materials. In this case,
the charges are bounded to the physical grains by the surface potential and
local superconductivity may arise in the interior. Long-range order or
supercurrents are a consequence of Josephson tunneling between the grains\cite{Ketterson}. 
Although the HTS crystals are not granular in a structural sense, the 
ubiquitous CDW in these materials led us to suggest\cite{DeMello2012} that they 
may form an array of mesoscopic Josephson junctions.

Under this assumption, the SC transition develops in two steps when the temperature 
decreases\cite{Ketterson}: Firstly, the order
parameters with local amplitudes $\Delta_d({\bf r}_i,p, T)$ and phases $\theta_i$ 
arise in each charge CDW domain $``i$''. 
{\color{red}
These localized amplitudes give rise to local Josephson coupling
$ E_{\rm J}(r_{ij}) $ that is proportional to th local supercurrent 
or th lattice version of the local superfluid 
density\cite{Spivak1991} $n_{\rm sc}$, and proportional to the local phase stiffness.}

Secondly, upon cooling more, 
the local phase stiffness increases and eventually overcomes
thermal disorder, which leads to a SC transition by long-range phase order.
Therefore, we emphasize that the SC critical temperature $T_{\rm c}$ 
is determined by the competition between
thermal disorder and the average planar Josephson energy
${\left  <E_{\rm J}(p,T) \right >}$.

These in-plane calculations are the fundamental pillars of the three-dimensional LRO
in the whole system, which we infer from transport measurements. 
For low doping $p$, the $c$-direction 
resistivity $\rho_c$ is $\approx 10^{3}-10^{6}$ larger 
than the $a$ or $b$-axis resistivity $\rho_{ab}$, a behavior shared also by 
${\mathrm{Bi}}_{2}{\mathrm{Sr}}_{2\ensuremath{-}x}{\mathrm{La}}_{x}{\mathrm{CuO}}_{6+\ensuremath{\delta}}$\cite{ResistBSLCO.2003,ResistC.ab.2004}.
Despite this huge difference, it is surprising that both $\rho_c(T)$, and $\rho_{ab}(T)$ fall 
to zero at the same temperature ($T_{\rm c}$). We have recently argued that the mechanism behind
this puzzling behavior may be understood in terms of the planar and weaker out-of-plane 
average Josephson coupling\cite{DeMello2012}, exactly like the weakly coupled XY models. 

As explained previously\cite{deMello2014}, even for $d$-wave amplitudes, 
it is sufficient to use the Ambegaokar-Baratoff analytical $s$-wave expression\cite{AB1963}
averaged over the plane:
\begin{equation}
 {\left < E_{\rm J}(p,T) \right >}^X = \frac{\pi \hbar {\left <\Delta_d(p,T)\right >}}
 {4 e^2 R^X_{\rm n}(p)} 
 {\rm tanh} \bigl [\frac{\left <\Delta_d(p,T)\right >}{2k_{\rm B}T} \bigr ] .
\label{EJ} 
\end{equation}
Where $X = ab$ for planar and $X = c$ for interlayer coupling and $R^X_{\rm n}(p, \sim T_{\rm c})$ 
is the corresponding normal state directional resistance 
just above $T_{\rm c}$. 
{\color{red}
In our model of an array of Josephson junctions, the current 
is composed of Cooper pairs tunneling between the CDW domains and by normal carriers or 
quasiparticle planar current\cite{Bruder95}. For a d-wave HTS near Tc the supercurrent is dominant\cite{Bruder95}, 
which justifies the use of the experimental $R_{\rm n} (T_{\rm c})$ between the charge domains in Eq. 6. 
}

As mentioned, thermal energy causes phase disorder and 
coherence is achieved\cite{DeMello2012,Mello2021} at 
${\left < E_{\rm J}(p,T_{\rm c}) \right >}^{ab} = k_{\rm B}T_{\rm c}$. 
The smaller planar resistances yield larger $E_{\rm J}^{ab}$ that promote first
LRO in the planes, but each plane ``$i$'' would have its own SC phase $\theta_i$
if it was not for the weaker inter-plane $E_{\rm J}^{c}$ coupling.
It is similar to a ferromagnet cooled down in the presence of a tiny
magnetic field causing all the moments to become aligned.

Thus, the weaker interlayer coupling ${\left < E_{\rm J}(p,T) \right >}^c$  
connects the planes but leads to only a single-phase $\theta$ at $T \le T_{\rm c}$ 
in the whole system, and both $c$ and $ab$ resistivity
drop off together despite their orders of magnitude difference.

\section{Magnetic Penetration depth and resistivity}

According to Eq. \ref{EJ}, due to the large difference in the directional resistivities,
we expect smaller superfluid densities $n_{\rm sc}$ along the 
$c$-direction than along the plane, which is confirmed by the $ab$ and $c$-axis 
penetration depth anisotropy\cite{Shibauchi94,Panagopoulos2000,Panagopoulos1999}. 
We recall also that the square of the magnetic penetration depth $\lambda$ is inversely proportional
to the  phase stiffness $\rho_{\rm sc}$\cite{Bozovic2016} that is 
proportional to the average Josephson 
current\cite{Spivak1991}. 

Along these lines and in the frame of the L-D model\cite{LD1971}, 
Shibauchi {\it et al}\cite{Shibauchi94} successfully reproduced their $c$-axis $\lambda_c(p)$
measurements. We extend here their approach to account 
for Josephson current between the CDW charge domains 
in the CuO planes and use that\cite{Mello2020a,Mello2021,Bozovic2016}
$\lambda^2_{X}(p) \propto 1/ {\left < E_{\rm J}(p,T) \right >}^{X}$.
Therefore, we may write:
\begin{equation}
 [\frac{ \lambda_{c}(0)}{\lambda_{ab}(0)}]^2 \propto [\frac{ {\left < E_{\rm J}(p,T) \right >}^{ab}} 
 {{\left < E_{\rm J}(p,T) \right >}^{c}} \propto \frac{ R^{c}_{\rm n}} {R^{ab}_{\rm n}},
 \end{equation}
and $ R^{c}_{\rm n} = \rho_c(T_{\rm c}) s$ 
where $s$ is the distance between the CuO layers in double-plane
LSCO crystals is approximately 6.6 $\AA$.
This means that $\lambda_c(p)$ is dominated by the interplane Josephson current
and, extending this idea, $\lambda_{ab}(p)$ is dominated
by the planar average coupling ${\left < E_{\rm J}(p,T_{\rm c}) \right >}^{ab}$. 
Therefore, we may write the planar resistivity
$ R^{ab}_{\rm n} = \rho_{ab}(T_{\rm c}) \lambda_{\rm CO}$, since $\lambda_{\rm CO}(p)$ 
is the distance between the planar CDW domains. Therefore,
in a general way,
\begin{equation}
 [\frac{ \lambda_{c}(0)}{\lambda_{ab}(0)}]^2 \propto \frac{ R^{c}_{\rm n}} {R^{ab}_{\rm n}}
 \propto  \frac{ \rho_c (T_{\rm c}) s} {\rho_{ab}(T_{\rm c}) \lambda_{\rm CO} }.
 \label{ratio}
 \end{equation}
This expression gives the magnetic penetration depths out of the plane 
and planar ratio in terms of similar resistivities ratio, the planar distance $s$, and
the CDW wavelength $\lambda_{\rm CO}$. Notice that {\it there is not any 
adjustable parameter in Eq. \ref{ratio}} and all quantities have previously been measured.

{\color{red} 
Some HTS samples with similar doping have comparable resistivities
like the La and Y-based compounds studied by Ando {\it et al}\cite{Ando2004}.
}
When this is the case
the above equation shows why the low temperature ratios $\lambda_{c}(p,0) / \lambda_{ab}(p,0)$ 
for different families of compounds
have similar values\cite{Schneider2004}. 
This is the case for a large number of 
LSCO and HgBa$_2$CuO$_{4+x}$ \cite{Panagopoulos2000} samples and they are also 
comparable with the measurements of $c$-axis grain-aligned orthorhombic 
YBa$_2$Cu$_3$O$_{7-\delta}$ (YBCO) with $\delta = 0.0, 0.3, $ and $ 0.43$\cite{Panagopoulos1998}.
The quantitative explanation of these data and their connection with 
the  $\rho_c(T_{\rm c})/ \rho_{ab}(T_{\rm c})$ 
is one of the main motivations of our present calculations.
\begin{figure}% figure* is for twocolumn figures.
%\begin{center}
%\centering
%\begin{minipage}{width=0.46\textwidth}
\centerline{\includegraphics[height=4.50cm]{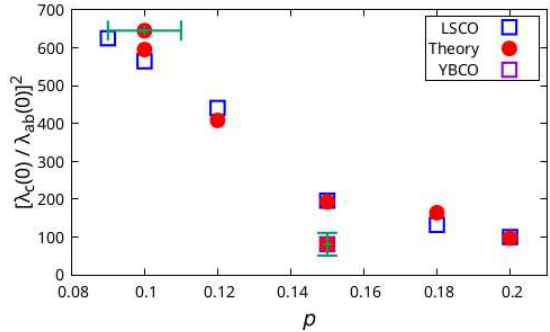}}
%end{center}
%i\end{minipage}
\caption{ Six LSCO $[\lambda_{c}(0)/\lambda_{ab}(0)]^2$ experimental 
points \cite{Shibauchi94,Panagopoulos2000} and one optimal YBCO-aligned powder in magnetic 
field with 30\% uncertainty\cite{Panagopoulos1998} represented by the error bar. They
are together with the calculations 
from Eq. \ref{ratio} which use the CDW wavelengths\cite{Comin2016} $\lambda_{CO}(p)$ 
and the respective resistivity ratios\cite{Kimura92,Nakamura93} 
listed in Tables \ref{table1} and \ref{table2}. For 
$p = 0.09$ we use the resistivities of
$p = 0.10 \pm 0.01$ from Ref. \onlinecite{Nakamura93} as explained in the text.
}
\label{fig2}
\end{figure}

On the other hand, most of the data on directional $\lambda_X(0)$ and $\rho_X(T_{\rm c})$ 
with the same doping $p$ were performed a long time ago with LSCO crystals in
order to understand the anisotropies in HTS. Nowadays there are single crystals of many
other materials but since the anisotropies are already established these
measurements are not remade.
The difficulty to fabricate single crystals in the earlier days of HTS 
is the reason why data on other materials are practically nonexistent.
In most cases, whenever there are data on $\lambda_{c}(0) / \lambda_{ab}(0)$, 
it is not accompanied by $\rho_c(T_{\rm c})/ \rho_{ab}(T_{\rm c})$ that is needed by
our Eq. \ref{ratio}. Nevertheless, we list in Table 1 the available
data on LSCO for $(\lambda_{c}(0)/\lambda_{ab}(0))^2$\cite{Shibauchi94,Panagopoulos2000} ratios and 
$\rho_c(T_{\rm c})/ \rho_{ab}(T_{\rm c})$\cite{Kimura92,Nakamura93}. 
The case of
$(\lambda_{c}(0.09)/\lambda_{ab}(0.09))^2$,  the resistivity data of different groups 
have discrepant results and we used $\rho_c/ \rho_{ab}(0.10 \pm 0.01) $ 
from Ref. \onlinecite{Nakamura93} (marked with a star).
With these data 
and the respective CDW  wavelengths $\lambda_{CO}(p)$ that enters in Eq. \ref{ratio} for 
the planar $ R^{ab}_{\rm n}$, we calculated the magnetic penetration depth ratio. 
The experimental results and our estimates are plotted together for comparison
in Fig. \ref{fig2}, and listed in columns two and five
of Table \ref{table1}.
%\begin{widetext}
\begin{table}[!ht]
\caption{ Data and calculations for LSCO. The first column is hole density per unit cell. 
Second is the 
$(\lambda_{c}(0)/\lambda_{ab}(0))^2$ measurements of Shibauchi {\it et al}\cite{Shibauchi94} and
Panagopoulos {\it et al}\cite{Panagopoulos2000}. Third is the 
$\rho_c(T_{\rm c})/ \rho_{ab}(T_{\rm c})$\cite{Kimura92,Nakamura93} resistivity ratios.  
The fourth is $\lambda_{CO}$ in units of the lattice parameter $a_0 \approx 3.78 \AA$
measured by REXS\cite{Comin2016}. The last column is the calculations
from Eq. \ref{ratio} with $s = 6.6  \AA$ that
should match the magnetic penetration depth ratio of the second column. The case of
$(\lambda_{c}(0.09)/\lambda_{ab}(0.09))^2$,  the resistivity data of different groups 
conflict and we used the $p = 0.10 \pm 0.01$ resistivity ratio $\rho_c/ \rho_{ab}$
 from Ref. \onlinecite{Nakamura93} (marked with a star).
}
 \begin{tabular}{|c|c|c|c|c|}\hline \hline
Sample & $(\lambda_{c}(0)/\lambda_{ab}(0))^2$ & $\rho_c/ \rho_{ab}(T_{\rm c})$
& $\lambda_{CO}(\AA)$ & $ \rho_c s/ \rho_{ab}\lambda_{CO} $  \\ \hline

 {\it p} = 0.09 & {\bf 625} &  2070* & 5.6 $a_0$ & {\bf 645}*\\ \hline
 {\it p} = 0.10 & {\bf $\propto$ 564} & 1714 &  5.0 $a_0$ & {\bf 595} \\ \hline
 {\it p} = 0.12 & {\bf 441} & 1000  & 4.25 $a_0$ & {\bf 408} \\ \hline
 {\it p} = 0.15 & {\bf 196}  & 433 & 3.9 $a_0$ & {\bf 193} \\ \hline
 {\it p} = 0.18 & {\bf 132} & 300 & 3.7 $a_0$ & {\bf 164} \\ \hline
 {\it p} = 0.20 & {\bf 100} & 200  & 3.6 $a_0$ & {\bf 97}\\ \hline
 \end{tabular}
 \label{table1}
% \end{center}
 \end{table}
 
There are also some data on $\lambda_{c}(0)/\lambda_{ab}(0)$ in magnetic 
aligned powder of YBa$_2$Cu$_3$O$_{7-\delta}$\cite{Panagopoulos1998} (YBCO) that contains some 
uncertainty up to 30\% in the $c$-axis but provides good estimates of this ratio.  
Similarly, we use 
new anisotropic $\rho_c(T_{\rm c})/ \rho_{ab}(T_{\rm c})$ optimal data on thin films of YBCO grown 
on an off-axis cut SrTiO3 substrate\cite{Yresist2021}. Combining these data, we can 
apply Eq. \ref{ratio} to this near optimal compound and the 
calculation is very close to the YBCO experimental result as shown in Fig. \ref{fig2} 
and in Table \ref{table2}.
\begin{table}[!ht]
\caption{ Data and calculations for optimal YBCO similar to Table 1. The uncertainty
of 30\% in $(\lambda_{c}(0)/\lambda_{ab}(0))^2$ is due to the $c$-axis alignment 
uncertainty of the powders. For YBCO, the 
distance between two planes is $s = 5.84 \AA$ and for optimal doping\cite{Comin2016}
$\lambda_{CO} = 3.12 a_0$.}
 \begin{tabular}{|c|c|c|c|c|}\hline \hline
Sample & $(\lambda_{c}(0)/\lambda_{ab}(0))^2$ & $\rho_c(T_{\rm c})/ \rho_{ab}(T_{\rm c})$
& $\lambda_{CO}(\AA)$ & $ \rho_c s/ \rho_{ab}\lambda_{CO} $ \\ \hline
{\it p} = 0.15 & {\bf 81 $\pm$ 25}  & 167 & 3.9 $a_0$ & {\bf 81.5} \\ \hline
\end{tabular}
 \label{table2}
% \end{center}
 \end{table}
 
 \section{Conclusion}
 
In this paper, we generalize the ideas of ILT and L-D models to account also for 
Josephson coupling between the mesoscopic CDW domains or blocks.
The in-plane average Josephson energy is much larger than the interlayer 
coupling, proportional to $1/ \lambda_{ab}^2$
and to the thermal energy at the critical temperature $T_{\rm c}$.
Therefore the SC properties are almost entirely dependent on the 
planar Josephson coupling in opposition to
the old ILT model, which depended on the interlayer Josephson tunneling. Our
approach yields values of $\lambda_{ab}$ at least one order of magnitude
larger than $\lambda_{c}$, which gives some insights
why measurements\cite{Images1998} of $\lambda_{c}$ gave much smaller condensation
energy than the old ILT predictions\cite{Leggett1996}. 

Furthermore, we derive a new equation relating the magnetic penetration depths
$(\lambda_{c}(0)/\lambda_{ab}(0))^2$ with the resistivities
$[\rho_c(T_{\rm c}) s]/ [\rho_{ab}(T_{\rm c}) \lambda_{\rm CO}]$ (Eq. \ref{ratio}), which 
is in agreement
with many measurements without any adjustable parameter. Nowadays there are
better single crystals of many
materials but since the anisotropies are already established these
combined measurements($\rho_{X}(T_{\rm c})$ and $(\lambda_{X}(0)$) on a 
single sample are not anymore explored. However, Eq. \ref{ratio} provides new motivation
for more precise tests in future experiments with modern pristine HTS crystals.

\section{acknowledgements}

We acknowledge partial support from the Brazilian agencies CNPq and FAPERJ.

%\bibliography{Anisot.Penet.bib}

\begin{thebibliography}{47}%
\makeatletter
\providecommand \@ifxundefined [1]{%
 \@ifx{#1\undefined}
}%
\providecommand \@ifnum [1]{%
 \ifnum #1\expandafter \@firstoftwo
 \else \expandafter \@secondoftwo
 \fi
}%
\providecommand \@ifx [1]{%
 \ifx #1\expandafter \@firstoftwo
 \else \expandafter \@secondoftwo
 \fi
}%
\providecommand \natexlab [1]{#1}%
\providecommand \enquote  [1]{``#1''}%
\providecommand \bibnamefont  [1]{#1}%
\providecommand \bibfnamefont [1]{#1}%
\providecommand \citenamefont [1]{#1}%
\providecommand \href@noop [0]{\@secondoftwo}%
\providecommand \href [0]{\begingroup \@sanitize@url \@href}%
\providecommand \@href[1]{\@@startlink{#1}\@@href}%
\providecommand \@@href[1]{\endgroup#1\@@endlink}%
\providecommand \@sanitize@url [0]{\catcode `\\12\catcode `\$12\catcode
  `\&12\catcode `\#12\catcode `\^12\catcode `\_12\catcode `\%12\relax}%
\providecommand \@@startlink[1]{}%
\providecommand \@@endlink[0]{}%
\providecommand \url  [0]{\begingroup\@sanitize@url \@url }%
\providecommand \@url [1]{\endgroup\@href {#1}{\urlprefix }}%
\providecommand \urlprefix  [0]{URL }%
\providecommand \Eprint [0]{\href }%
\providecommand \doibase [0]{http://dx.doi.org/}%
\providecommand \selectlanguage [0]{\@gobble}%
\providecommand \bibinfo  [0]{\@secondoftwo}%
\providecommand \bibfield  [0]{\@secondoftwo}%
\providecommand \translation [1]{[#1]}%
\providecommand \BibitemOpen [0]{}%
\providecommand \bibitemStop [0]{}%
\providecommand \bibitemNoStop [0]{.\EOS\space}%
\providecommand \EOS [0]{\spacefactor3000\relax}%
\providecommand \BibitemShut  [1]{\csname bibitem#1\endcsname}%
\let\auto@bib@innerbib\@empty
%</preamble>
\bibitem [{\citenamefont {Clem}(1989)}]{Clem89}%
  \BibitemOpen
  \bibfield  {author} {\bibinfo {author} {\bibfnamefont {J.~R.}\ \bibnamefont
  {Clem}},\ }\href {\doibase https://doi.org/10.1016/0921-4534(89)90630-8}
  {\bibfield  {journal} {\bibinfo  {journal} {Physica C}\ }\textbf {\bibinfo
  {volume} {162-164}},\ \bibinfo {pages} {1137} (\bibinfo {year}
  {1989})}\BibitemShut {NoStop}%
\bibitem [{\citenamefont {Clem}(1991)}]{Clem91}%
  \BibitemOpen
  \bibfield  {author} {\bibinfo {author} {\bibfnamefont {J.~R.}\ \bibnamefont
  {Clem}},\ }\href {\doibase 10.1103/PhysRevB.43.7837} {\bibfield  {journal}
  {\bibinfo  {journal} {Phys. Rev. B}\ }\textbf {\bibinfo {volume} {43}},\
  \bibinfo {pages} {7837} (\bibinfo {year} {1991})}\BibitemShut {NoStop}%
\bibitem [{\citenamefont {Chakravarty}\ \emph {et~al.}(1993)\citenamefont
  {Chakravarty}, \citenamefont {Sudbo}, \citenamefont {Anderson},\ and\
  \citenamefont {Strong}}]{Chakravarty1993}%
  \BibitemOpen
  \bibfield  {author} {\bibinfo {author} {\bibfnamefont {S.}~\bibnamefont
  {Chakravarty}}, \bibinfo {author} {\bibfnamefont {A.}~\bibnamefont {Sudbo}},
  \bibinfo {author} {\bibfnamefont {P.~W.}\ \bibnamefont {Anderson}}, \ and\
  \bibinfo {author} {\bibfnamefont {S.}~\bibnamefont {Strong}},\ }\href
  {https://link.gale.com/apps/doc/A14360657/AONE?u=uff_br&sid=googleScholar&xid=2ff9e9b6}
  {\bibfield  {journal} {\bibinfo  {journal} {Science}\ ,\ \bibinfo {pages}
  {337+}} (\bibinfo {year} {1993})},\ \bibinfo {note} {5119}\BibitemShut
  {NoStop}%
\bibitem [{\citenamefont {Anderson}(1995)}]{Anderson1995}%
  \BibitemOpen
  \bibfield  {author} {\bibinfo {author} {\bibfnamefont {P.}~\bibnamefont
  {Anderson}},\ }\href
  {https://link.gale.com/apps/doc/A17029914/AONE?u=uff_br&sid=googleScholar&xid=7c61657d}
  {\bibfield  {journal} {\bibinfo  {journal} {Science}\ }\textbf {\bibinfo
  {volume} {268}},\ \bibinfo {pages} {1154} (\bibinfo {year}
  {1995})}\BibitemShut {NoStop}%
\bibitem [{\citenamefont {Leggett}(1996)}]{Leggett1996}%
  \BibitemOpen
  \bibfield  {author} {\bibinfo {author} {\bibfnamefont {A.~J.}\ \bibnamefont
  {Leggett}},\ }\href
  {https://www.science.org/doi/abs/10.1126/science.274.5287.587} {\bibfield
  {journal} {\bibinfo  {journal} {Science}\ }\textbf {\bibinfo {volume}
  {274}},\ \bibinfo {pages} {587} (\bibinfo {year} {1996})}\BibitemShut
  {NoStop}%
\bibitem [{\citenamefont {Lawrence}\ and\ \citenamefont
  {Doniach}(1971)}]{LD1971}%
  \BibitemOpen
  \bibfield  {author} {\bibinfo {author} {\bibfnamefont {W.}~\bibnamefont
  {Lawrence}}\ and\ \bibinfo {author} {\bibfnamefont {S.}~\bibnamefont
  {Doniach}},\ }in\ \href@noop {} {\emph {\bibinfo {booktitle} {Proceedings of
  the Twelfth International Conference on Low Temperature Physics}}}\ (\bibinfo
   {publisher} {Ed. by E. Kanda, Academic Press of Japan},\ \bibinfo {address}
  {Kyoto, Japan},\ \bibinfo {year} {1971})\ p.\ \bibinfo {pages}
  {361}\BibitemShut {NoStop}%
\bibitem [{\citenamefont {Shibauchi}\ \emph {et~al.}(1994)\citenamefont
  {Shibauchi}, \citenamefont {Kitano}, \citenamefont {Uchinokura},
  \citenamefont {Maeda}, \citenamefont {Kimura},\ and\ \citenamefont
  {Kishio}}]{Shibauchi94}%
  \BibitemOpen
  \bibfield  {author} {\bibinfo {author} {\bibfnamefont {T.}~\bibnamefont
  {Shibauchi}}, \bibinfo {author} {\bibfnamefont {H.}~\bibnamefont {Kitano}},
  \bibinfo {author} {\bibfnamefont {K.}~\bibnamefont {Uchinokura}}, \bibinfo
  {author} {\bibfnamefont {A.}~\bibnamefont {Maeda}}, \bibinfo {author}
  {\bibfnamefont {T.}~\bibnamefont {Kimura}}, \ and\ \bibinfo {author}
  {\bibfnamefont {K.}~\bibnamefont {Kishio}},\ }\href {\doibase
  10.1103/PhysRevLett.72.2263} {\bibfield  {journal} {\bibinfo  {journal}
  {Phys. Rev. Lett.}\ }\textbf {\bibinfo {volume} {72}},\ \bibinfo {pages}
  {2263} (\bibinfo {year} {1994})}\BibitemShut {NoStop}%
\bibitem [{\citenamefont {Moler}\ \emph {et~al.}(1998)\citenamefont {Moler},
  \citenamefont {Kirtley}, \citenamefont {Hinks}, \citenamefont {Li},\ and\
  \citenamefont {Xu}}]{Images1998}%
  \BibitemOpen
  \bibfield  {author} {\bibinfo {author} {\bibfnamefont {K.~A.}\ \bibnamefont
  {Moler}}, \bibinfo {author} {\bibfnamefont {J.~R.}\ \bibnamefont {Kirtley}},
  \bibinfo {author} {\bibfnamefont {D.}~\bibnamefont {Hinks}}, \bibinfo
  {author} {\bibfnamefont {T.}~\bibnamefont {Li}}, \ and\ \bibinfo {author}
  {\bibfnamefont {M.}~\bibnamefont {Xu}},\ }\href
  {https://link.gale.com/apps/doc/A20355173/AONE?u=uff_br&sid=googleScholar&xid=f12940d3}
  {\bibfield  {journal} {\bibinfo  {journal} {Science}\ }\textbf {\bibinfo
  {volume} {279}},\ \bibinfo {pages} {1193} (\bibinfo {year}
  {1998})}\BibitemShut {NoStop}%
\bibitem [{\citenamefont {Anderson}(1998)}]{Anderson1998}%
  \BibitemOpen
  \bibfield  {author} {\bibinfo {author} {\bibfnamefont {P.~W.}\ \bibnamefont
  {Anderson}},\ }\href
  {https://link.gale.com/apps/doc/A20355174/AONE?u=uff_br&sid=googleScholar&xid=b14e3a35}
  {\bibfield  {journal} {\bibinfo  {journal} {Science}\ }\textbf {\bibinfo
  {volume} {279}},\ \bibinfo {pages} {1196+} (\bibinfo {year} {1998})},\
  \bibinfo {note} {5354}\BibitemShut {NoStop}%
\bibitem [{\citenamefont {Tranquada}\ \emph {et~al.}(1995)\citenamefont
  {Tranquada}, \citenamefont {Sternlieb}, \citenamefont {Axe}, \citenamefont
  {Nakamura},\ and\ \citenamefont {Uchida}}]{Tranquada1995a}%
  \BibitemOpen
  \bibfield  {author} {\bibinfo {author} {\bibfnamefont {J.~M.}\ \bibnamefont
  {Tranquada}}, \bibinfo {author} {\bibfnamefont {B.~J.}\ \bibnamefont
  {Sternlieb}}, \bibinfo {author} {\bibfnamefont {J.~D.}\ \bibnamefont {Axe}},
  \bibinfo {author} {\bibfnamefont {Y.}~\bibnamefont {Nakamura}}, \ and\
  \bibinfo {author} {\bibfnamefont {S.}~\bibnamefont {Uchida}},\ }\href
  {\doibase 10.1038/375561a0} {\bibfield  {journal} {\bibinfo  {journal}
  {Nature}\ }\textbf {\bibinfo {volume} {375}},\ \bibinfo {pages} {561}
  (\bibinfo {year} {1995})}\BibitemShut {NoStop}%
\bibitem [{\citenamefont {Lang}\ \emph {et~al.}(2002)\citenamefont {Lang},
  \citenamefont {Madhavan}, \citenamefont {Hoffman}, \citenamefont {Hudson},
  \citenamefont {Eisaki}, \citenamefont {Uchida},\ and\ \citenamefont
  {Davis}}]{Lang2002}%
  \BibitemOpen
  \bibfield  {author} {\bibinfo {author} {\bibfnamefont {K.~M.}\ \bibnamefont
  {Lang}}, \bibinfo {author} {\bibfnamefont {V.}~\bibnamefont {Madhavan}},
  \bibinfo {author} {\bibfnamefont {J.~E.}\ \bibnamefont {Hoffman}}, \bibinfo
  {author} {\bibfnamefont {E.~W.}\ \bibnamefont {Hudson}}, \bibinfo {author}
  {\bibfnamefont {H.}~\bibnamefont {Eisaki}}, \bibinfo {author} {\bibfnamefont
  {S.}~\bibnamefont {Uchida}}, \ and\ \bibinfo {author} {\bibfnamefont {J.~C.}\
  \bibnamefont {Davis}},\ }\href {\doibase 10.1038/415412a} {\bibfield
  {journal} {\bibinfo  {journal} {Nature}\ }\textbf {\bibinfo {volume} {415}},\
  \bibinfo {pages} {412} (\bibinfo {year} {2002})}\BibitemShut {NoStop}%
\bibitem [{\citenamefont {Fischer}\ \emph {et~al.}(2007)\citenamefont
  {Fischer}, \citenamefont {Kugler}, \citenamefont {Maggio-Aprile},
  \citenamefont {Berthod},\ and\ \citenamefont {Renner}}]{STM2007}%
  \BibitemOpen
  \bibfield  {author} {\bibinfo {author} {\bibfnamefont {O.}~\bibnamefont
  {Fischer}}, \bibinfo {author} {\bibfnamefont {M.}~\bibnamefont {Kugler}},
  \bibinfo {author} {\bibfnamefont {I.}~\bibnamefont {Maggio-Aprile}}, \bibinfo
  {author} {\bibfnamefont {C.}~\bibnamefont {Berthod}}, \ and\ \bibinfo
  {author} {\bibfnamefont {C.}~\bibnamefont {Renner}},\ }\href {\doibase
  10.1103/RevModPhys.79.353} {\bibfield  {journal} {\bibinfo  {journal} {Rev.
  Mod. Phys.}\ }\textbf {\bibinfo {volume} {79}},\ \bibinfo {pages} {353}
  (\bibinfo {year} {2007})}\BibitemShut {NoStop}%
\bibitem [{\citenamefont {Comin}\ and\ \citenamefont
  {Damascelli}(2016)}]{Comin2016}%
  \BibitemOpen
  \bibfield  {author} {\bibinfo {author} {\bibfnamefont {R.}~\bibnamefont
  {Comin}}\ and\ \bibinfo {author} {\bibfnamefont {A.}~\bibnamefont
  {Damascelli}},\ }\href {\doibase 10.1146/annurev-conmatphys-031115-011401}
  {\bibfield  {journal} {\bibinfo  {journal} {Ann. Rev. of Cond. Mat. Phys.}\
  }\textbf {\bibinfo {volume} {7}},\ \bibinfo {pages} {369} (\bibinfo {year}
  {2016})}\BibitemShut {NoStop}%
\bibitem [{\citenamefont {Wise}\ \emph {et~al.}(2008)\citenamefont {Wise} \emph
  {et~al.}}]{Wise2008}%
  \BibitemOpen
  \bibfield  {author} {\bibinfo {author} {\bibfnamefont {W.~D.}\ \bibnamefont
  {Wise}} \emph {et~al.},\ }\href {\doibase 10.1038/nphys1021} {\bibfield
  {journal} {\bibinfo  {journal} {Nature Physics}\ }\textbf {\bibinfo {volume}
  {4}},\ \bibinfo {pages} {696} (\bibinfo {year} {2008})}\BibitemShut {NoStop}%
\bibitem [{\citenamefont {Gomes}\ \emph {et~al.}(2007)\citenamefont {Gomes}
  \emph {et~al.}}]{Gomes2007}%
  \BibitemOpen
  \bibfield  {author} {\bibinfo {author} {\bibfnamefont {K.~K.}\ \bibnamefont
  {Gomes}} \emph {et~al.},\ }\href {\doibase 10.1038/nature05881} {\bibfield
  {journal} {\bibinfo  {journal} {Nature}\ }\textbf {\bibinfo {volume} {447}},\
  \bibinfo {pages} {569} (\bibinfo {year} {2007})}\BibitemShut {NoStop}%
\bibitem [{\citenamefont {Parker}\ \emph {et~al.}(2010)\citenamefont {Parker},
  \citenamefont {Aynajian}, \citenamefont {{da Silva Neto}}, \citenamefont
  {Pushp}, \citenamefont {Ono}, \citenamefont {Wen}, \citenamefont {Xu},
  \citenamefont {Gu},\ and\ \citenamefont {Yazdani}}]{Parker2010}%
  \BibitemOpen
  \bibfield  {author} {\bibinfo {author} {\bibfnamefont {C.~V.}\ \bibnamefont
  {Parker}}, \bibinfo {author} {\bibfnamefont {P.}~\bibnamefont {Aynajian}},
  \bibinfo {author} {\bibfnamefont {E.~H.}\ \bibnamefont {{da Silva Neto}}},
  \bibinfo {author} {\bibfnamefont {A.}~\bibnamefont {Pushp}}, \bibinfo
  {author} {\bibfnamefont {S.}~\bibnamefont {Ono}}, \bibinfo {author}
  {\bibfnamefont {J.}~\bibnamefont {Wen}}, \bibinfo {author} {\bibfnamefont
  {Z.}~\bibnamefont {Xu}}, \bibinfo {author} {\bibfnamefont {G.}~\bibnamefont
  {Gu}}, \ and\ \bibinfo {author} {\bibfnamefont {A.}~\bibnamefont {Yazdani}},\
  }\href {http://dx.doi.org/10.1038/nature09597} {\bibfield  {journal}
  {\bibinfo  {journal} {Nature}\ }\textbf {\bibinfo {volume} {468}},\ \bibinfo
  {pages} {677} (\bibinfo {year} {2010})}\BibitemShut {NoStop}%
\bibitem [{\citenamefont {He}\ \emph {et~al.}(2014)\citenamefont {He},
  \citenamefont {Yin}, \citenamefont {Zech}, \citenamefont {Soumyanarayanan},
  \citenamefont {Yee}, \citenamefont {Williams}, \citenamefont {Boyer},
  \citenamefont {Chatterjee}, \citenamefont {Wise}, \citenamefont {Zeljkovic},
  \citenamefont {Kondo}, \citenamefont {Takeuchi}, \citenamefont {Ikuta},
  \citenamefont {Mistark}, \citenamefont {Markiewicz}, \citenamefont {Bansil},
  \citenamefont {Sachdev}, \citenamefont {Hudson},\ and\ \citenamefont
  {Hoffman}}]{He2014}%
  \BibitemOpen
  \bibfield  {author} {\bibinfo {author} {\bibfnamefont {Y.}~\bibnamefont
  {He}}, \bibinfo {author} {\bibfnamefont {Y.}~\bibnamefont {Yin}}, \bibinfo
  {author} {\bibfnamefont {M.}~\bibnamefont {Zech}}, \bibinfo {author}
  {\bibfnamefont {A.}~\bibnamefont {Soumyanarayanan}}, \bibinfo {author}
  {\bibfnamefont {M.~M.}\ \bibnamefont {Yee}}, \bibinfo {author} {\bibfnamefont
  {T.}~\bibnamefont {Williams}}, \bibinfo {author} {\bibfnamefont {M.~C.}\
  \bibnamefont {Boyer}}, \bibinfo {author} {\bibfnamefont {K.}~\bibnamefont
  {Chatterjee}}, \bibinfo {author} {\bibfnamefont {W.~D.}\ \bibnamefont
  {Wise}}, \bibinfo {author} {\bibfnamefont {I.}~\bibnamefont {Zeljkovic}},
  \bibinfo {author} {\bibfnamefont {T.}~\bibnamefont {Kondo}}, \bibinfo
  {author} {\bibfnamefont {T.}~\bibnamefont {Takeuchi}}, \bibinfo {author}
  {\bibfnamefont {H.}~\bibnamefont {Ikuta}}, \bibinfo {author} {\bibfnamefont
  {P.}~\bibnamefont {Mistark}}, \bibinfo {author} {\bibfnamefont {R.~S.}\
  \bibnamefont {Markiewicz}}, \bibinfo {author} {\bibfnamefont
  {A.}~\bibnamefont {Bansil}}, \bibinfo {author} {\bibfnamefont
  {S.}~\bibnamefont {Sachdev}}, \bibinfo {author} {\bibfnamefont {E.~W.}\
  \bibnamefont {Hudson}}, \ and\ \bibinfo {author} {\bibfnamefont {J.~E.}\
  \bibnamefont {Hoffman}},\ }\href {\doibase 10.1126/science.1248221}
  {\bibfield  {journal} {\bibinfo  {journal} {Science}\ }\textbf {\bibinfo
  {volume} {344}},\ \bibinfo {pages} {608} (\bibinfo {year}
  {2014})}\BibitemShut {NoStop}%
\bibitem [{\citenamefont {da~Silva~Neto}\ \emph {et~al.}()\citenamefont
  {da~Silva~Neto}, \citenamefont {Minola}, \citenamefont {Yu}, \citenamefont
  {Tabis}, \citenamefont {Bluschke}, \citenamefont {Unruh}, \citenamefont
  {Suzuki}, \citenamefont {Li}, \citenamefont {Yu}, \citenamefont {Betto},
  \citenamefont {Kummer}, \citenamefont {Yakhou}, \citenamefont {Brookes},
  \citenamefont {Le~Tacon}, \citenamefont {Greven}, \citenamefont {Keimer},\
  and\ \citenamefont {Damascelli}}]{Ndoped2018}%
  \BibitemOpen
  \bibfield  {author} {\bibinfo {author} {\bibfnamefont {E.~H.}\ \bibnamefont
  {da~Silva~Neto}}, \bibinfo {author} {\bibfnamefont {M.}~\bibnamefont
  {Minola}}, \bibinfo {author} {\bibfnamefont {B.}~\bibnamefont {Yu}}, \bibinfo
  {author} {\bibfnamefont {W.}~\bibnamefont {Tabis}}, \bibinfo {author}
  {\bibfnamefont {M.}~\bibnamefont {Bluschke}}, \bibinfo {author}
  {\bibfnamefont {D.}~\bibnamefont {Unruh}}, \bibinfo {author} {\bibfnamefont
  {H.}~\bibnamefont {Suzuki}}, \bibinfo {author} {\bibfnamefont
  {Y.}~\bibnamefont {Li}}, \bibinfo {author} {\bibfnamefont {G.}~\bibnamefont
  {Yu}}, \bibinfo {author} {\bibfnamefont {D.}~\bibnamefont {Betto}}, \bibinfo
  {author} {\bibfnamefont {K.}~\bibnamefont {Kummer}}, \bibinfo {author}
  {\bibfnamefont {F.}~\bibnamefont {Yakhou}}, \bibinfo {author} {\bibfnamefont
  {N.~B.}\ \bibnamefont {Brookes}}, \bibinfo {author} {\bibfnamefont
  {M.}~\bibnamefont {Le~Tacon}}, \bibinfo {author} {\bibfnamefont
  {M.}~\bibnamefont {Greven}}, \bibinfo {author} {\bibfnamefont
  {B.}~\bibnamefont {Keimer}}, \ and\ \bibinfo {author} {\bibfnamefont
  {A.}~\bibnamefont {Damascelli}},\ }\href@noop {} {\ }\BibitemShut {NoStop}%
\bibitem [{\citenamefont {Wu}\ \emph {et~al.}(2017)\citenamefont {Wu},
  \citenamefont {Bollinger}, \citenamefont {He},\ and\ \citenamefont
  {Bo{\v{z}}ovi{\'c}}}]{Wu2017}%
  \BibitemOpen
  \bibfield  {author} {\bibinfo {author} {\bibfnamefont {J.}~\bibnamefont
  {Wu}}, \bibinfo {author} {\bibfnamefont {A.~T.}\ \bibnamefont {Bollinger}},
  \bibinfo {author} {\bibfnamefont {X.}~\bibnamefont {He}}, \ and\ \bibinfo
  {author} {\bibfnamefont {I.}~\bibnamefont {Bo{\v{z}}ovi{\'c}}},\ }\href
  {http://dx.doi.org/10.1038/nature23290} {\bibfield  {journal} {\bibinfo
  {journal} {Nature}\ }\textbf {\bibinfo {volume} {547}},\ \bibinfo {pages}
  {432} (\bibinfo {year} {2017})}\BibitemShut {NoStop}%
\bibitem [{\citenamefont {Chen}\ \emph {et~al.}(2019)\citenamefont {Chen},
  \citenamefont {Hashimoto}, \citenamefont {He}, \citenamefont {Song},
  \citenamefont {Xu}, \citenamefont {He}, \citenamefont {Devereaux},
  \citenamefont {Eisaki}, \citenamefont {Lu}, \citenamefont {Zaanen},\ and\
  \citenamefont {Shen}}]{Shen2019}%
  \BibitemOpen
  \bibfield  {author} {\bibinfo {author} {\bibfnamefont {S.-D.}\ \bibnamefont
  {Chen}}, \bibinfo {author} {\bibfnamefont {M.}~\bibnamefont {Hashimoto}},
  \bibinfo {author} {\bibfnamefont {Y.}~\bibnamefont {He}}, \bibinfo {author}
  {\bibfnamefont {D.}~\bibnamefont {Song}}, \bibinfo {author} {\bibfnamefont
  {K.-J.}\ \bibnamefont {Xu}}, \bibinfo {author} {\bibfnamefont {J.-F.}\
  \bibnamefont {He}}, \bibinfo {author} {\bibfnamefont {T.~P.}\ \bibnamefont
  {Devereaux}}, \bibinfo {author} {\bibfnamefont {H.}~\bibnamefont {Eisaki}},
  \bibinfo {author} {\bibfnamefont {D.-H.}\ \bibnamefont {Lu}}, \bibinfo
  {author} {\bibfnamefont {J.}~\bibnamefont {Zaanen}}, \ and\ \bibinfo {author}
  {\bibfnamefont {Z.-X.}\ \bibnamefont {Shen}},\ }\href {\doibase
  10.1126/science.aaw8850} {\bibfield  {journal} {\bibinfo  {journal}
  {Science}\ }\textbf {\bibinfo {volume} {366}},\ \bibinfo {pages} {1099}
  (\bibinfo {year} {2019})}\BibitemShut {NoStop}%
\bibitem [{\citenamefont {Fei}\ \emph {et~al.}(2019)\citenamefont {Fei},
  \citenamefont {Zheng}, \citenamefont {Bu}, \citenamefont {Zhang},
  \citenamefont {Ding}, \citenamefont {Zhou},\ and\ \citenamefont
  {Yin}}]{Fei2019}%
  \BibitemOpen
  \bibfield  {author} {\bibinfo {author} {\bibfnamefont {Y.}~\bibnamefont
  {Fei}}, \bibinfo {author} {\bibfnamefont {Y.}~\bibnamefont {Zheng}}, \bibinfo
  {author} {\bibfnamefont {K.}~\bibnamefont {Bu}}, \bibinfo {author}
  {\bibfnamefont {W.}~\bibnamefont {Zhang}}, \bibinfo {author} {\bibfnamefont
  {Y.}~\bibnamefont {Ding}}, \bibinfo {author} {\bibfnamefont {X.}~\bibnamefont
  {Zhou}}, \ and\ \bibinfo {author} {\bibfnamefont {Y.}~\bibnamefont {Yin}},\
  }\href {\doibase 10.1007/s11433-019-9454-6} {\bibfield  {journal} {\bibinfo
  {journal} {Science China Physics, Mechanics \& Astronomy}\ }\textbf {\bibinfo
  {volume} {63}},\ \bibinfo {pages} {227411} (\bibinfo {year}
  {2019})}\BibitemShut {NoStop}%
\bibitem [{\citenamefont {Miao}\ \emph {et~al.}(2021)\citenamefont {Miao},
  \citenamefont {Fabbris}, \citenamefont {Koch}, \citenamefont {Mazzone},
  \citenamefont {Nelson}, \citenamefont {Acevedo-Esteves}, \citenamefont {Gu},
  \citenamefont {Li}, \citenamefont {Yilimaz}, \citenamefont {Kaznatcheev},
  \citenamefont {Vescovo}, \citenamefont {Oda}, \citenamefont {Kurosawa},
  \citenamefont {Momono}, \citenamefont {Assefa}, \citenamefont {Robinson},
  \citenamefont {Bozin}, \citenamefont {Tranquada}, \citenamefont {Johnson},\
  and\ \citenamefont {Dean}}]{Tranquada2021}%
  \BibitemOpen
  \bibfield  {author} {\bibinfo {author} {\bibfnamefont {H.}~\bibnamefont
  {Miao}}, \bibinfo {author} {\bibfnamefont {G.}~\bibnamefont {Fabbris}},
  \bibinfo {author} {\bibfnamefont {R.~J.}\ \bibnamefont {Koch}}, \bibinfo
  {author} {\bibfnamefont {D.~G.}\ \bibnamefont {Mazzone}}, \bibinfo {author}
  {\bibfnamefont {C.~S.}\ \bibnamefont {Nelson}}, \bibinfo {author}
  {\bibfnamefont {R.}~\bibnamefont {Acevedo-Esteves}}, \bibinfo {author}
  {\bibfnamefont {G.~D.}\ \bibnamefont {Gu}}, \bibinfo {author} {\bibfnamefont
  {Y.}~\bibnamefont {Li}}, \bibinfo {author} {\bibfnamefont {T.}~\bibnamefont
  {Yilimaz}}, \bibinfo {author} {\bibfnamefont {K.}~\bibnamefont
  {Kaznatcheev}}, \bibinfo {author} {\bibfnamefont {E.}~\bibnamefont
  {Vescovo}}, \bibinfo {author} {\bibfnamefont {M.}~\bibnamefont {Oda}},
  \bibinfo {author} {\bibfnamefont {T.}~\bibnamefont {Kurosawa}}, \bibinfo
  {author} {\bibfnamefont {N.}~\bibnamefont {Momono}}, \bibinfo {author}
  {\bibfnamefont {T.}~\bibnamefont {Assefa}}, \bibinfo {author} {\bibfnamefont
  {I.~K.}\ \bibnamefont {Robinson}}, \bibinfo {author} {\bibfnamefont {E.~S.}\
  \bibnamefont {Bozin}}, \bibinfo {author} {\bibfnamefont {J.~M.}\ \bibnamefont
  {Tranquada}}, \bibinfo {author} {\bibfnamefont {P.~D.}\ \bibnamefont
  {Johnson}}, \ and\ \bibinfo {author} {\bibfnamefont {M.~P.~M.}\ \bibnamefont
  {Dean}},\ }\href {\doibase 10.1038/s41535-021-00327-4} {\bibfield  {journal}
  {\bibinfo  {journal} {npj Quantum Materials}\ }\textbf {\bibinfo {volume}
  {6}},\ \bibinfo {pages} {31} (\bibinfo {year} {2021})}\BibitemShut {NoStop}%
\bibitem [{\citenamefont {de~Mello}(2012)}]{DeMello2012}%
  \BibitemOpen
  \bibfield  {author} {\bibinfo {author} {\bibfnamefont {E.~V.~L.}\
  \bibnamefont {de~Mello}},\ }\href {\doibase 10.1209/0295-5075/99/37003}
  {\bibfield  {journal} {\bibinfo  {journal} {Europhys. Lett.}\ }\textbf
  {\bibinfo {volume} {99}},\ \bibinfo {pages} {37003} (\bibinfo {year}
  {2012})}\BibitemShut {NoStop}%
\bibitem [{\citenamefont {de~Mello}\ and\ \citenamefont
  {Sonier}(2014)}]{deMello2014}%
  \BibitemOpen
  \bibfield  {author} {\bibinfo {author} {\bibfnamefont {E.~V.~L.}\
  \bibnamefont {de~Mello}}\ and\ \bibinfo {author} {\bibfnamefont {J.~E.}\
  \bibnamefont {Sonier}},\ }\href
  {http://stacks.iop.org/0953-8984/26/i=49/a=492201} {\bibfield  {journal}
  {\bibinfo  {journal} {J. Phys.: Condens. Matter}\ }\textbf {\bibinfo {volume}
  {26}},\ \bibinfo {pages} {492201} (\bibinfo {year} {2014})}\BibitemShut
  {NoStop}%
\bibitem [{\citenamefont {de~Mello}\ and\ \citenamefont
  {Sonier}(2017)}]{Mello2017}%
  \BibitemOpen
  \bibfield  {author} {\bibinfo {author} {\bibfnamefont {E.~V.~L.}\
  \bibnamefont {de~Mello}}\ and\ \bibinfo {author} {\bibfnamefont {J.~E.}\
  \bibnamefont {Sonier}},\ }\href {\doibase 10.1103/PhysRevB.95.184520}
  {\bibfield  {journal} {\bibinfo  {journal} {Phys. Rev. B}\ }\textbf {\bibinfo
  {volume} {95}},\ \bibinfo {pages} {184520} (\bibinfo {year}
  {2017})}\BibitemShut {NoStop}%
\bibitem [{\citenamefont {de~Mello}(2020)}]{Mello2020a}%
  \BibitemOpen
  \bibfield  {author} {\bibinfo {author} {\bibfnamefont {E.~V.}\ \bibnamefont
  {de~Mello}},\ }\href {\doibase 10.1088/1361-648x/ab9fd5} {\bibfield
  {journal} {\bibinfo  {journal} {J. Phys.: Condens. Matter}\ }\textbf
  {\bibinfo {volume} {32}},\ \bibinfo {pages} {40LT02} (\bibinfo {year}
  {2020})}\BibitemShut {NoStop}%
\bibitem [{\citenamefont {Santana}\ and\ \citenamefont
  {de~Mello}(2022)}]{Mello2022}%
  \BibitemOpen
  \bibfield  {author} {\bibinfo {author} {\bibfnamefont {H.~S.}\ \bibnamefont
  {Santana}}\ and\ \bibinfo {author} {\bibfnamefont {E.}~\bibnamefont
  {de~Mello}},\ }\href {\doibase 10.1103/PhysRevB.105.134513} {\bibfield
  {journal} {\bibinfo  {journal} {Phys. Rev. B}\ }\textbf {\bibinfo {volume}
  {105}},\ \bibinfo {pages} {134513} (\bibinfo {year} {2022})}\BibitemShut
  {NoStop}%
\bibitem [{\citenamefont {Spivak}\ and\ \citenamefont
  {Kivelson}(1991)}]{Spivak1991}%
  \BibitemOpen
  \bibfield  {author} {\bibinfo {author} {\bibfnamefont {B.~I.}\ \bibnamefont
  {Spivak}}\ and\ \bibinfo {author} {\bibfnamefont {S.~A.}\ \bibnamefont
  {Kivelson}},\ }\href {\doibase 10.1103/PhysRevB.43.3740} {\bibfield
  {journal} {\bibinfo  {journal} {Phys. Rev. B}\ }\textbf {\bibinfo {volume}
  {43}},\ \bibinfo {pages} {3740} (\bibinfo {year} {1991})}\BibitemShut
  {NoStop}%
\bibitem [{\citenamefont {Bo{\v{z}}ovi{\'c}}\ \emph {et~al.}(2016)\citenamefont
  {Bo{\v{z}}ovi{\'c}}, \citenamefont {He}, \citenamefont {Wu},\ and\
  \citenamefont {Bollinger}}]{Bozovic2016}%
  \BibitemOpen
  \bibfield  {author} {\bibinfo {author} {\bibfnamefont {I.}~\bibnamefont
  {Bo{\v{z}}ovi{\'c}}}, \bibinfo {author} {\bibfnamefont {X.}~\bibnamefont
  {He}}, \bibinfo {author} {\bibfnamefont {J.}~\bibnamefont {Wu}}, \ and\
  \bibinfo {author} {\bibfnamefont {A.~T.}\ \bibnamefont {Bollinger}},\ }\href
  {http://dx.doi.org/10.1038/nature19061} {\bibfield  {journal} {\bibinfo
  {journal} {Nature}\ }\textbf {\bibinfo {volume} {536}},\ \bibinfo {pages}
  {309} (\bibinfo {year} {2016})}\BibitemShut {NoStop}%
\bibitem [{\citenamefont {Panagopoulos}\ \emph {et~al.}(2000)\citenamefont
  {Panagopoulos}, \citenamefont {Cooper}, \citenamefont {Xiang}, \citenamefont
  {Wang},\ and\ \citenamefont {Chu}}]{Panagopoulos2000}%
  \BibitemOpen
  \bibfield  {author} {\bibinfo {author} {\bibfnamefont {C.}~\bibnamefont
  {Panagopoulos}}, \bibinfo {author} {\bibfnamefont {J.~R.}\ \bibnamefont
  {Cooper}}, \bibinfo {author} {\bibfnamefont {T.}~\bibnamefont {Xiang}},
  \bibinfo {author} {\bibfnamefont {Y.~S.}\ \bibnamefont {Wang}}, \ and\
  \bibinfo {author} {\bibfnamefont {C.~W.}\ \bibnamefont {Chu}},\ }\href
  {\doibase 10.1103/PhysRevB.61.R3808} {\bibfield  {journal} {\bibinfo
  {journal} {Phys. Rev. B}\ }\textbf {\bibinfo {volume} {61}},\ \bibinfo
  {pages} {R3808} (\bibinfo {year} {2000})}\BibitemShut {NoStop}%
\bibitem [{\citenamefont {Kimura}\ \emph {et~al.}(1992)\citenamefont {Kimura},
  \citenamefont {Kishio}, \citenamefont {Kobayashi}, \citenamefont {Nakayama},
  \citenamefont {Motohira}, \citenamefont {Kitazawa},\ and\ \citenamefont
  {Yamafuji}}]{Kimura92}%
  \BibitemOpen
  \bibfield  {author} {\bibinfo {author} {\bibfnamefont {T.}~\bibnamefont
  {Kimura}}, \bibinfo {author} {\bibfnamefont {K.}~\bibnamefont {Kishio}},
  \bibinfo {author} {\bibfnamefont {T.}~\bibnamefont {Kobayashi}}, \bibinfo
  {author} {\bibfnamefont {Y.}~\bibnamefont {Nakayama}}, \bibinfo {author}
  {\bibfnamefont {N.}~\bibnamefont {Motohira}}, \bibinfo {author}
  {\bibfnamefont {K.}~\bibnamefont {Kitazawa}}, \ and\ \bibinfo {author}
  {\bibfnamefont {K.}~\bibnamefont {Yamafuji}},\ }\href {\doibase
  https://doi.org/10.1016/0921-4534(92)90767-7} {\bibfield  {journal} {\bibinfo
   {journal} {Physica C: Superconductivity}\ }\textbf {\bibinfo {volume}
  {192}},\ \bibinfo {pages} {247} (\bibinfo {year} {1992})}\BibitemShut
  {NoStop}%
\bibitem [{\citenamefont {Nakamura}\ and\ \citenamefont
  {Uchida}(1993)}]{Nakamura93}%
  \BibitemOpen
  \bibfield  {author} {\bibinfo {author} {\bibfnamefont {Y.}~\bibnamefont
  {Nakamura}}\ and\ \bibinfo {author} {\bibfnamefont {S.}~\bibnamefont
  {Uchida}},\ }\href {\doibase 10.1103/PhysRevB.47.8369} {\bibfield  {journal}
  {\bibinfo  {journal} {Phys. Rev. B}\ }\textbf {\bibinfo {volume} {47}},\
  \bibinfo {pages} {8369} (\bibinfo {year} {1993})}\BibitemShut {NoStop}%
\bibitem [{\citenamefont {de~Mello}\ and\ \citenamefont
  {da~Silveira~Filho}(2005)}]{Otton2005}%
  \BibitemOpen
  \bibfield  {author} {\bibinfo {author} {\bibfnamefont {E.}~\bibnamefont
  {de~Mello}}\ and\ \bibinfo {author} {\bibfnamefont {O.~T.}\ \bibnamefont
  {da~Silveira~Filho}},\ }\href {\doibase
  http://dx.doi.org/10.1016/j.physa.2004.08.076} {\bibfield  {journal}
  {\bibinfo  {journal} {Physica A}\ }\textbf {\bibinfo {volume} {347}},\
  \bibinfo {pages} {429 } (\bibinfo {year} {2005})}\BibitemShut {NoStop}%
\bibitem [{\citenamefont {de~Mello}\ \emph {et~al.}(2009)\citenamefont
  {de~Mello}, \citenamefont {Kasal},\ and\ \citenamefont
  {Passos}}]{deMello2009}%
  \BibitemOpen
  \bibfield  {author} {\bibinfo {author} {\bibfnamefont {E.~V.~L.}\
  \bibnamefont {de~Mello}}, \bibinfo {author} {\bibfnamefont {R.~B.}\
  \bibnamefont {Kasal}}, \ and\ \bibinfo {author} {\bibfnamefont {C.~A.~C.}\
  \bibnamefont {Passos}},\ }\href
  {http://stacks.iop.org/0953-8984/21/i=23/a=235701} {\bibfield  {journal}
  {\bibinfo  {journal} {J. Phys.: Condens. Matter}\ }\textbf {\bibinfo {volume}
  {21}},\ \bibinfo {pages} {235701} (\bibinfo {year} {2009})}\BibitemShut
  {NoStop}%
\bibitem [{\citenamefont {Bray}(1994)}]{Bray1994}%
  \BibitemOpen
  \bibfield  {author} {\bibinfo {author} {\bibfnamefont {A.}~\bibnamefont
  {Bray}},\ }\href {\doibase 10.1080/00018739400101505} {\bibfield  {journal}
  {\bibinfo  {journal} {Adv. Phys.}\ }\textbf {\bibinfo {volume} {43}},\
  \bibinfo {pages} {357} (\bibinfo {year} {1994})}\BibitemShut {NoStop}%
\bibitem [{\citenamefont {de~Mello}(2021)}]{Mello2021}%
  \BibitemOpen
  \bibfield  {author} {\bibinfo {author} {\bibfnamefont {E.~V.~L.}\
  \bibnamefont {de~Mello}},\ }\href {\doibase 10.1088/1361-648x/abd812}
  {\bibfield  {journal} {\bibinfo  {journal} {J. of Phys.: Cond. Matter}\
  }\textbf {\bibinfo {volume} {33}},\ \bibinfo {pages} {145503} (\bibinfo
  {year} {2021})}\BibitemShut {NoStop}%
\bibitem [{\citenamefont {Chang}\ \emph {et~al.}(2012)\citenamefont {Chang},
  \citenamefont {Blackburn}, \citenamefont {Holmes}, \citenamefont
  {Christensen}, \citenamefont {Larsen}, \citenamefont {Mesot}, \citenamefont
  {Liang}, \citenamefont {Bonn}, \citenamefont {Hardy}, \citenamefont
  {Watenphul}, \citenamefont {Zimmermann}, \citenamefont {Forgan},\ and\
  \citenamefont {Hayden}}]{Chang2012}%
  \BibitemOpen
  \bibfield  {author} {\bibinfo {author} {\bibfnamefont {J.}~\bibnamefont
  {Chang}}, \bibinfo {author} {\bibfnamefont {E.}~\bibnamefont {Blackburn}},
  \bibinfo {author} {\bibfnamefont {T.}~\bibnamefont {Holmes}}, \bibinfo
  {author} {\bibfnamefont {N.~B.}\ \bibnamefont {Christensen}}, \bibinfo
  {author} {\bibfnamefont {J.}~\bibnamefont {Larsen}}, \bibinfo {author}
  {\bibfnamefont {J.}~\bibnamefont {Mesot}}, \bibinfo {author} {\bibfnamefont
  {R.}~\bibnamefont {Liang}}, \bibinfo {author} {\bibfnamefont {D.~A.}\
  \bibnamefont {Bonn}}, \bibinfo {author} {\bibfnamefont {W.~N.}\ \bibnamefont
  {Hardy}}, \bibinfo {author} {\bibfnamefont {A.}~\bibnamefont {Watenphul}},
  \bibinfo {author} {\bibfnamefont {M.~V.}\ \bibnamefont {Zimmermann}},
  \bibinfo {author} {\bibfnamefont {E.~M.}\ \bibnamefont {Forgan}}, \ and\
  \bibinfo {author} {\bibfnamefont {S.~M.}\ \bibnamefont {Hayden}},\ }\href
  {\doibase 10.1038/nphys2456} {\bibfield  {journal} {\bibinfo  {journal}
  {Nature Physics}\ }\textbf {\bibinfo {volume} {8}},\ \bibinfo {pages} {871}
  (\bibinfo {year} {2012})}\BibitemShut {NoStop}%
\bibitem [{\citenamefont {Ketterson}\ and\ \citenamefont
  {Song}(1999)}]{Ketterson}%
  \BibitemOpen
  \bibfield  {author} {\bibinfo {author} {\bibfnamefont {J.~B.}\ \bibnamefont
  {Ketterson}}\ and\ \bibinfo {author} {\bibfnamefont {S.}~\bibnamefont
  {Song}},\ }\href@noop {} {\emph {\bibinfo {title} {Superconductivity}}}\
  (\bibinfo  {publisher} {Cambridge University Press},\ \bibinfo {address}
  {London},\ \bibinfo {year} {1999})\BibitemShut {NoStop}%
\bibitem [{\citenamefont {Ono}\ and\ \citenamefont
  {Ando}(2003)}]{ResistBSLCO.2003}%
  \BibitemOpen
  \bibfield  {author} {\bibinfo {author} {\bibfnamefont {S.}~\bibnamefont
  {Ono}}\ and\ \bibinfo {author} {\bibfnamefont {Y.}~\bibnamefont {Ando}},\
  }\href {\doibase 10.1103/PhysRevB.67.104512} {\bibfield  {journal} {\bibinfo
  {journal} {Phys. Rev. B}\ }\textbf {\bibinfo {volume} {67}},\ \bibinfo
  {pages} {104512} (\bibinfo {year} {2003})}\BibitemShut {NoStop}%
\bibitem [{\citenamefont {Komiya}\ \emph {et~al.}(2002)\citenamefont {Komiya},
  \citenamefont {Ando}, \citenamefont {Sun},\ and\ \citenamefont
  {Lavrov}}]{ResistC.ab.2004}%
  \BibitemOpen
  \bibfield  {author} {\bibinfo {author} {\bibfnamefont {S.}~\bibnamefont
  {Komiya}}, \bibinfo {author} {\bibfnamefont {Y.}~\bibnamefont {Ando}},
  \bibinfo {author} {\bibfnamefont {X.~F.}\ \bibnamefont {Sun}}, \ and\
  \bibinfo {author} {\bibfnamefont {A.~N.}\ \bibnamefont {Lavrov}},\ }\href
  {\doibase 10.1103/PhysRevB.65.214535} {\bibfield  {journal} {\bibinfo
  {journal} {Phys. Rev. B}\ }\textbf {\bibinfo {volume} {65}},\ \bibinfo
  {pages} {214535} (\bibinfo {year} {2002})}\BibitemShut {NoStop}%
\bibitem [{\citenamefont {Ambegaokar}\ and\ \citenamefont
  {Baratoff}(1963)}]{AB1963}%
  \BibitemOpen
  \bibfield  {author} {\bibinfo {author} {\bibfnamefont {V.}~\bibnamefont
  {Ambegaokar}}\ and\ \bibinfo {author} {\bibfnamefont {A.}~\bibnamefont
  {Baratoff}},\ }\href {\doibase 10.1103/PhysRevLett.10.486} {\bibfield
  {journal} {\bibinfo  {journal} {Phys. Rev. Lett.}\ }\textbf {\bibinfo
  {volume} {10}},\ \bibinfo {pages} {486} (\bibinfo {year} {1963})}\BibitemShut
  {NoStop}%
\bibitem [{\citenamefont {Bruder}\ \emph {et~al.}(1995)\citenamefont {Bruder},
  \citenamefont {van Otterlo},\ and\ \citenamefont {Zimanyi}}]{Bruder95}%
  \BibitemOpen
  \bibfield  {author} {\bibinfo {author} {\bibfnamefont {C.}~\bibnamefont
  {Bruder}}, \bibinfo {author} {\bibfnamefont {A.}~\bibnamefont {van Otterlo}},
  \ and\ \bibinfo {author} {\bibfnamefont {G.~T.}\ \bibnamefont {Zimanyi}},\
  }\href {\doibase 10.1103/PhysRevB.51.12904} {\bibfield  {journal} {\bibinfo
  {journal} {Phys. Rev. B}\ }\textbf {\bibinfo {volume} {51}},\ \bibinfo
  {pages} {12904} (\bibinfo {year} {1995})}\BibitemShut {NoStop}%
\bibitem [{\citenamefont {Panagopoulos}\ \emph {et~al.}(1999)\citenamefont
  {Panagopoulos}, \citenamefont {Tallon},\ and\ \citenamefont
  {Xiang}}]{Panagopoulos1999}%
  \BibitemOpen
  \bibfield  {author} {\bibinfo {author} {\bibfnamefont {C.}~\bibnamefont
  {Panagopoulos}}, \bibinfo {author} {\bibfnamefont {J.~L.}\ \bibnamefont
  {Tallon}}, \ and\ \bibinfo {author} {\bibfnamefont {T.}~\bibnamefont
  {Xiang}},\ }\href {\doibase 10.1103/PhysRevB.59.R6635} {\bibfield  {journal}
  {\bibinfo  {journal} {Phys. Rev. B}\ }\textbf {\bibinfo {volume} {59}},\
  \bibinfo {pages} {R6635} (\bibinfo {year} {1999})}\BibitemShut {NoStop}%
\bibitem [{\citenamefont {Ando}\ \emph {et~al.}(2004)\citenamefont {Ando},
  \citenamefont {Komiya}, \citenamefont {Segawa}, \citenamefont {Ono},\ and\
  \citenamefont {Kurita}}]{Ando2004}%
  \BibitemOpen
  \bibfield  {author} {\bibinfo {author} {\bibfnamefont {Y.}~\bibnamefont
  {Ando}}, \bibinfo {author} {\bibfnamefont {S.}~\bibnamefont {Komiya}},
  \bibinfo {author} {\bibfnamefont {K.}~\bibnamefont {Segawa}}, \bibinfo
  {author} {\bibfnamefont {S.}~\bibnamefont {Ono}}, \ and\ \bibinfo {author}
  {\bibfnamefont {Y.}~\bibnamefont {Kurita}},\ }\href {\doibase
  10.1103/PhysRevLett.93.267001} {\bibfield  {journal} {\bibinfo  {journal}
  {Phys. Rev. Lett.}\ }\textbf {\bibinfo {volume} {93}},\ \bibinfo {pages}
  {267001} (\bibinfo {year} {2004})}\BibitemShut {NoStop}%
\bibitem [{\citenamefont {Schneider}\ and\ \citenamefont
  {Keller}(2004)}]{Schneider2004}%
  \BibitemOpen
  \bibfield  {author} {\bibinfo {author} {\bibfnamefont {T.}~\bibnamefont
  {Schneider}}\ and\ \bibinfo {author} {\bibfnamefont {H.}~\bibnamefont
  {Keller}},\ }\href {\doibase 10.1088/1367-2630/6/1/144} {\bibfield  {journal}
  {\bibinfo  {journal} {New Journal of Physics}\ }\textbf {\bibinfo {volume}
  {6}},\ \bibinfo {pages} {144} (\bibinfo {year} {2004})}\BibitemShut {NoStop}%
\bibitem [{\citenamefont {Panagopoulos}\ \emph {et~al.}(1998)\citenamefont
  {Panagopoulos}, \citenamefont {Cooper},\ and\ \citenamefont
  {Xiang}}]{Panagopoulos1998}%
  \BibitemOpen
  \bibfield  {author} {\bibinfo {author} {\bibfnamefont {C.}~\bibnamefont
  {Panagopoulos}}, \bibinfo {author} {\bibfnamefont {J.~R.}\ \bibnamefont
  {Cooper}}, \ and\ \bibinfo {author} {\bibfnamefont {T.}~\bibnamefont
  {Xiang}},\ }\href {\doibase 10.1103/PhysRevB.57.13422} {\bibfield  {journal}
  {\bibinfo  {journal} {Phys. Rev. B}\ }\textbf {\bibinfo {volume} {57}},\
  \bibinfo {pages} {13422} (\bibinfo {year} {1998})}\BibitemShut {NoStop}%
\bibitem [{\citenamefont {Heine}\ \emph {et~al.}(2021)\citenamefont {Heine},
  \citenamefont {Lang}, \citenamefont {Rössler},\ and\ \citenamefont
  {Pedarnig}}]{Yresist2021}%
  \BibitemOpen
  \bibfield  {author} {\bibinfo {author} {\bibfnamefont {G.}~\bibnamefont
  {Heine}}, \bibinfo {author} {\bibfnamefont {W.}~\bibnamefont {Lang}},
  \bibinfo {author} {\bibfnamefont {R.}~\bibnamefont {Rössler}}, \ and\
  \bibinfo {author} {\bibfnamefont {J.~D.}\ \bibnamefont {Pedarnig}},\ }\href
  {\doibase 10.3390/nano11030675} {\bibfield  {journal} {\bibinfo  {journal}
  {Nanomaterials}\ }\textbf {\bibinfo {volume} {11}} (\bibinfo {year} {2021}),\
  10.3390/nano11030675}\BibitemShut {NoStop}%
\end{thebibliography}
%
\end{document}